\documentstyle[12pt]{article}

\begin{document}
\input{psfig.sty}
\begin{flushright}
\baselineskip=12pt
MADPH-99-1137 \\
\end{flushright}

\begin{center}
\vglue 1.5cm
{\Large\bf  Gauge Hierarchy from $AdS_5$ Universe with 3-Branes\\}
\vglue 2.0cm
{\Large Tianjun Li~\footnote{E-mail: li@pheno.physics.wisc.edu,
phone: (608) 262-9820, fax: (608) 262-8628.}}
\vglue 1cm
\begin{flushleft}
Department of Physics, University of Wisconsin, Madison, WI 53706,  
U.  S.  A.
\end{flushleft}
\end{center}

\vglue 1.5cm
\begin{abstract}
If the universe is ( or a slice of ) $AdS_5$ space with 3-branes,
 the 5-dimensional GUT scale on 
each brane
can be indetified as the 5-dimensional Planck scale, but,
the 4-dimensional Planck scale is generated from the low 4-dimensional
GUT scale exponentially in our world.
 The 4-dimensional GUT scales and Planck scale
are related to the 5-dimensional GUT scales and Planck scale 
by exponential factors, respectively.
 One of such scenarios was
suggested by Randall and Sundrum recently. We give another scenario that
the 4-dimensional Planck scale is
generated from the low five-dimensional Planck scale by
an exponential hierarchy, and the mass scale in the Standard Model
is not rescaled from the 5-dimensional metric to the
4-dimensional metric.      
We also argue that the additional constant in the 
solution might exist, which  will rescale the 5-dimensional Planck scale
and affect the physical scale picture. 
Finally, we embed 
those compactifications to the general compactification
on $AdS_5$ space and discuss the origin of the additional constant.  
\\[1ex]
\\[1ex]

\end{abstract}

\vspace{0.5cm}
\begin{flushleft}
\baselineskip=12pt
August 1999\\
\end{flushleft}
\newpage
\setcounter{page}{1}
\pagestyle{plain}
\baselineskip=14pt

\section{Introduction}

Experiments at LEP and Tevatron have given the strong
support to the Standard Model of the Strong and electroweak 
interactions. However, the Standard Model has some unattractive features which
may imply the new physics beyond Standard Model. 
One of these problems is that the gauge forces and the gravitational force
are not unified. Another is
the gauge hierarchy problem between the weak scale and the 4-dimensional
Planck scale. Previously, two solutions to the
gauge hierarchy problem have
been proposed: one is the idea of the technicolor and  compositeness
which lack calculability, and the other is the idea
of supersymmetry, but we have not found any experimental signal
at colliders yet.     

More than one year ago, it was suggested that the large 
compactified extra dimensions may be another solution to the
gauge hierarchy problem~\cite{AADD}. If the dimension of the spacetime is 
$4+n$ where $n > 1$, the 4-dimensional Planck scale is determined
by the fundamental $4+n$ dimensional Planck scale $M_X$. For the
simplest case, we have the relation between these two scales:
\begin{eqnarray}
M_{Pl}^2 = M_X^{n+2} V_n  
~,~\,
\end{eqnarray}
where $V_n$ is the physical volume of the extra dimensions.
Of course, if we required that $M_X$ is at low energy scale, then 
we may have the hierarchy between the $M_X$ and $V_n^{-1/n}$
except that one needs to introduce many extra dimensions. If 
we assume $M_X=10^5 GeV$, $V_n^{-1/n} = 10^4 GeV$, we obtain:
\begin{eqnarray}
n = 10^{26}
~,~\,
\end{eqnarray}
which might not seem possible, but 
it is not excluded.
 
Several months ago, Randall and Sundrum~\cite{LRRS} proposed another scenario
that the extra dimension is an orbifold,
and  the size of the extra dimension is not large
but the 4-dimensional mass scale in the standard model is
suppressed by an exponential factor from 5-dimensional mass
scale. In addition,  they suggested that
the fifth dimension might be infinity~\cite{LRRSN}, and there may exist only one
brane with positive tension at the origin, but, there  exists a gauge
hierarchy problem for they thought $k$,  which is defined in the following
section, is
equal to the 5-dimensional Planck scale.  The remarkable aspect of
the second scenario is that it gives rise to a localized graviton field.
Combining those results, Lykken and Randall obtained the following
physical picture~\cite{JLLR}: the graviton is localized on Planck brane, we
live on a brane separated from the Planck brane about 30 Planck lengths 
along the fifth dimension. On our brane, the mass scale
in the Standard Model is suppressed
exponentially, which gives the low energy scale.
Recently, this kind of compactification or similar idea has attracted
much attentions~\cite{MG, HDDK, MBW, TN, CGKT, NK, HV, IO, ABKS, AK}.

Before discussing our results, first, we would like to point out
that  one needs to choose
a fundamental metric in the scale discussion. For example, in the
weakly coupled string, to be consistent,
 we  use the string metric to discuss the scale.  In addition,
 we assume that all the gauge forces are unified on each 3-brane
 if there exist gauge forces. The 5-dimensional GUT scale
 on each 3-brane and the 5-dimensional Planck scale are defined as
 the GUT scale and Planck scale in the 5-dimensional fundamental
 metric ( for example, equation (3) or (4) ), respectively. 
 The 4-dimensional GUT scale on each brane and the 4-dimenisonal
 Planck scale are defined as the GUT scale and Planck scale in the
 4-dimensional Minkowski Metric ($g_{\mu \nu}$). In order to
 avoid the gauge hierarchy problem between the weak scale and
 the 4-dimensional GUT scale $M_{GUT}$ on the brane which is
 our world, we assume the low energy unification~\footnote{We 
 will not explain why $M_{GUT}$ can be low energy scale here,
but, it is possible if one considers additional particles which 
change the RGE running. Of course, proton decay might be the
problem, but we do not disscuss this here.}. The
 key ansatz is that the 5-dimensional GUT scale on each brane 
 is equal to the 5-dimensional Planck scale, and there is
 no mass hierarchy between the mass parameters in the 
 5-dimensional fundamental metric  if the  fifth dimension is compact
\footnote{We do not consider the masses of quarks and leptons in this 
statement.}.

With above argument and assumptions, we find out the constant in the solution
of 5-dimensional Einstein equation is not a trivial factor, and
will affect the 5-dimensional Planck scale and 
the physical picture if there exists such constant in
the fundamental metric, for example, in M-thoery on $S^1/Z_2$\cite{HW}, 
only the 
metric at mid point between two hyperplanes is not changed after one
considers the next order correction without five branes
~\cite{OVRUT}, and
the constant in the metric may have its origin in the compactification
on $AdS_5$ space.
In short, the general solution to the 
Einstein equation  obtained in 
five-dimensional model with two boundaries (branes)
are:
\begin{eqnarray}
ds^2 = e^{- 2 k |y| -2 c} \eta_{\mu \nu} dx^{\mu} dx^{\nu} + d y^2 
~,~\,
\end{eqnarray}
\begin{eqnarray}
ds^2 = e^{ 2 k |y| -2 c} \eta_{\mu \nu} dx^{\mu} dx^{\nu} + d y^2 
~,~\,
\end{eqnarray}  
where $\mu, \nu$ parametrize the four-dimensional coordinates of
Minkowski space, and y is  the fifth dimension coordinate.

The first solution is the same as the solution given by 
Randall and Sundrum except the constant c~\cite{LRRS}. 
 The general physical scale picture is the following: 
 the 5-dimensional GUT scale on each brane
can be identified as the 5-dimensional Planck scale, but,
the 4-dimensional Planck scale is generated from the low 4-dimensional
GUT scale exponentially in our world.
 The 4-dimensional GUT scales and Planck scale
are related to the 5-dimensional GUT scales and Planck scale 
by exponential factors, respectively. In other words,
the large mass hierarchy~\cite{LRRS} in the 
Standard Model from the 5-dimensional metric
to the 4-dimensional metric is  one possible solution
of the gauge hierarchy problem in our world, but, not the most
general solution.
In addition,  the fifth dimension may be compact~\cite{LRRS}, 
or the fifth dimension may be 
non-compact. There is a brane with positive tension ( call it as
`` Hidden Brane'' ) at the origin, the 4-dimensional GUT scale $M_H$ on 
the hidden
brane  is  larger than  or the same as 
the 4-dimensional GUT scale of our world.   We
live on a brane separated from the hidden 
 brane about a distance $ln(M_H/M_{GUT})/k$ 
along the fifth dimension where $k$ 
 is not necessary to be equal to the 5-dimensional Planck scale  in
this case for the fifth dimension is not compact.
Similar to Ref. [3], we also suggest that: we live on
a positive tension brane ( at origin ) which is embedded in a non-compact
$AdS_5$ space and contains the Standard Model. The five-dimensional Planck scale
or the 4-dimensional GUT scale is at intermediate scale $10^{8}$ GeV,
then, in this case, $M_{GUT}/M_W \simeq M_{top}/M_{electron}$.

The second solution tells us that
the 5-dimensional Planck scale will be rescaled even though
one considers $c=0$. Assuming c=0, we can
obtain the scenario with the following property: 
the 4-dimensional Planck scale is
generated from the low 5-dimensional Planck scale by
an exponential hierachy, and the mass scale in the Standard Model, 
which is 
contained in one brane, is not rescaled. 
In general, for $c\neq 0$,  the 5-dimensional GUT scale on each brane
can be considered as the 5-dimensional Planck scale, but,
the 4-dimensional Planck scale is generated from the low 4-dimensional
GUT scale exponentially in our world.
The  4-dimensional GUT scales and Planck scale
are related to the 5-dimensional GUT scales and Planck scale 
by exponential factors, respectively. The physical scale picture is similar 
to that in the
first solution
except that we can not let the fifth dimension non-compact in this
case.

We also discuss the compactification on $AdS_5$ space, and all of
above scenarios can be embedded in this kind of compactification.
However, in all of above scenarios, the  cosmology constant
and brane tensions in the 4-dimensional Minkowski metric
can be expressed as  simple functions of 4-dimensional
GUT scale and Planck scale, therefore, it seems to us that we can not
explain the very small cosmology constant $10^{-58} GeV^4$ in this
kind of compactification.  

Of course, the results in this letter can be generalized to 
$AdS_{4+n}$, and can be applied to Type IIB string on $AdS_5\times S^5$
or M-theory on
$AdS_7\times S^4$.

\section{Solutions to the Einstein Equation}
The setup is given by Randall and Sundrum~\cite{LRRS}. Considering a compact 
fifth dimension, $ -L \leq y \leq L$ and introducing  equivalence 
relations: $ y \sim y+ 2L$ and
$ y \sim -y$, we obtain the obifold $S^1/Z_2$. The 
fixed points y=0, L  are taken as the locations of the two 3-branes,
which can support 4-dimensional field theories. Let us
denote the y=0, L planes as $M_0$ and $M_1$ respectively,
the 5-dimensional metric in these two branes are:
\begin{equation}
\label{smmetric}
g_{\mu \nu}^{0}(x^{\mu}) \equiv G_{\mu \nu}(x^{\mu}, y=0) ~,~
g_{\mu \nu}^{1}(x^{\mu}) \equiv G_{\mu \nu}(x^{\mu}, y=L) ~,~\,
\end{equation}
where $G_{AB}$ where $A, B = \mu, y$, is the five-dimensional metric.

The classical Lagrangian is given by:

\begin{eqnarray}
S &=& S_{gravity} + S_{0} + S_{1} 
~,~\,
\end{eqnarray}
\begin{eqnarray}
S_{gravity} &=& 
\int d^4 x \int_{- L}^{L} dy \sqrt{-G} \{- \Lambda + 
{1\over 2} M_X^3 R \} 
~,~\,
\end{eqnarray}
\begin{eqnarray}
S_{0} &=& \int d^4 x \sqrt{-g^0} \{ {\cal L}_{0} 
-  V_{0} \} 
~,~\,
\end{eqnarray}
\begin{eqnarray}
S_{1} &=& \int d^4 x \sqrt{-g^1} \{ {\cal L}_{1} -  V_{1} \}
~,~\,
\end{eqnarray} 
where $M_X$ is the 5-dimensional Planck scale,
$\Lambda$ is the cosmology constant, and 
$V_0$, $V_1$ are the brane tensions.
The 5-dimensional Einstein equation for the  above action is~\cite{LRRS}:
\begin{eqnarray} 
\sqrt{-G} \left( R_{AB}-{1 \over 2 } G_{AB} R \right) &=& - \frac{1}{ M_X^3} 
[ \Lambda \sqrt{-G} ~G_{AB} +  V_{0} \sqrt{-g^{0}} ~g_{\mu \nu}^{0} 
~\delta^\mu_M \delta^\nu_N ~\delta(y) \nonumber \\ 
&+& 
 V_{1} \sqrt{-g^{1}}  ~g_{\mu \nu}^{1} 
~\delta^\mu_M \delta^\nu_N ~\delta(y-L) ] ~.~ \,
\end{eqnarray}
Assuming that there exists a solution that
 respects   4-dimensional 
Poincare invariance in the $x^{\mu}$-directions, one obtains
the 5-dimensional metric:
\begin{eqnarray} 
ds^2 = e^{- 2 \sigma(y)} \eta_{\mu \nu} dx^{\mu} dx^{\nu}
 + dy^2 ~.~\, 
\end{eqnarray}
With this metric, the Einstein equation reduces to~\cite{LRRS}:
\begin{eqnarray}
\sigma^{\prime 2} = { - {\Lambda} \over { 6 M_X^3}} ~,~\, 
\end{eqnarray}
\begin{eqnarray}
 \sigma^{\prime \prime} =  
{{V_{0}} \over\displaystyle {3 M_X^3 } } \delta (y) +
{{V_{1}}\over\displaystyle {3 M_X^3}} \delta (y-L) 
~.~\, 
\end{eqnarray}

Before we discuss the phenomenology, we
would like to  define the generalized scale transformation
~\footnote{ Conformal invariance is a consequence of the scale invariance
in quantum field theory.}
on the 4-dimensional Lagrangian:
\begin{equation}
g_{\mu \nu} \rightarrow \lambda^2 g_{\mu \nu} ~,~
\phi \rightarrow \lambda^{-1} \phi ~,~\,
\end{equation} 
\begin{equation}
\psi \rightarrow \lambda^{-3/2} \psi ~,~
A_{\mu} \rightarrow A_{\mu} ~,~\,
\end{equation} 
\begin{equation}
m \rightarrow \lambda^{-1} m ~,~
g_c \rightarrow \lambda^{-dim[g_c]} ~,~ \, 
\end{equation} 
where $\phi$, $\psi$, $A_{\mu}$, $m$ and $g_c$ are 
the scalar field, spinor field, gauge field, mass and
coupling, respectively, and $dim[g_c]$ is the mass dimension
of the coupling ( for example, the gauge coupling and Yukawa 
coupling have dimension 0 ).
Obviously, if $\lambda$ is not a function of the space-time index,
one can easily check that
the generalized scale transformation is invariant even when one
considers the gravity, the divergence and renormalization. It
does not change the relative mass ratios,
 for example, if we consider the  Standard
Model and  gravity in 4-dimension, $M_W/M_{pl}$ is 
invariant under this
transformation. 
Although we do not mention it explicitly,
this kind of the transformation will be often used
in the following discussion because one need to consider the
physical mass scale in the  
 4-dimensional Minkowski metric
  from the 5-dimensional fundamental
 metric by the compactification of the fifth dimension. 
 By the way, we can define the generalized scale transformation
 in any dimension.

Now, we consider the solution.

(I)  Assuming $\Lambda$ is negative, the first
solution is:
\begin{equation}
\sigma =  |y| \sqrt{{{- {\Lambda}} \over\displaystyle {6 M_X^3}}} + c
~.~\,
\end{equation}
This solution was obtained by Randall and Sundrum by 
choosing c=0~\cite{LRRS}. Similar to their result, one
can define a new scale k which relates $V_0, V_1, \Lambda$ to $M_X$ as:
\begin{equation}
V_{0} = - V_{1} = 6 M_X^3 k~,~ \Lambda = - 6 M_X^3 k^2
~.~\,
\end{equation} 
Then, the bulk metric is:
\begin{equation}
ds^2 = e^{- 2 k  |y| - 2 c} \eta_{\mu \nu} dx^{\mu} dx^{\nu} + dy^2
~.~ \,
\end{equation}
The corresponding 4-dimensional Planck scale  is:
\begin{equation}
M_{pl}^2 =  {{M_X^3  e^{-2c}} \over\displaystyle k} [1- e^{-2 k L}]
 ~.~ \,
\end{equation} 
We assume that the observable brane is $M_1$,
 and in the 5-dimensional fundamental metric, the GUT scale
 $M_{GUT}^{(5)}$ of our world is equal to
the Planck scale $M_X$. We obtain our world 4-dimensional GUT scale
\begin{equation}  
M_{GUT} = M_{GUT}^{(5)} e^{-k L - c} = M_X e^{-k L - c} ~,~ \,
\end{equation} 
and the 4-dimensional Planck scale and GUT scale
relation
\begin{equation}
M_{pl}^2 =  {{M_{GUT}^3  e^{+3 k L + c}} \over\displaystyle k} [1- e^{-2 k L}]
 ~.~\,
\end{equation}
We can push $M_{GUT}$ to the TeV scale or $10^5$ GeV
easily. Assuming that in the fundamental theory we just have one scale, 
$L^{-1}$ should be close to the $M_X$. Because $ k L$  can not be arbitrarily 
large~\footnote{For $kL$ is small, the physical
results are similar to those in previous extra dimension
proposal~\cite{AADD},  
 so, we will not discuss it here.},
 we choose $k=M_X$ for the simplicity and obtain:
\begin{equation}
M_{pl}^2 =  M_{GUT}^2  e^{2 k L } [1- e^{-2 k L}]
 ~.~\,
\end{equation}  
With $k L= 34.5$ and 30, we can have  $M_{GUT}$ at
the TeV scale  and $10^5$ GeV scale respectively.
For c=0 and $k=M_X$, we obtain $M_X=M_{pl}$. For
$c=-kL/2$ and $k=M_X$, we obtain that for $M_{GUT} = 10^3$ GeV and $10^5$ GeV,
$M_X$ is about $10^{10}$ GeV and $ 10^{11}$ GeV, respectively.
And for $c= kL$, we obtain $ M_X = M_{GUT}$. In short, for $c \neq 0$,
the physical scale picture is different from that in Ref. [2,3].  
And the result in Ref.[2] is not the most general solution
to the gauge hierarchy problem in our world.  

One can also consider the cosmology constant and brane tensions in
the 4-dimensional Minkowski metric, they are:
\begin{eqnarray}
\Lambda^{(4)}=-{{3 M_{GUT}^6 e^{6kL}} \over\displaystyle 
M_{pl}^2} (1-e^{-2kL}) (1-e^{-4kL}) 
~,~\,
\end{eqnarray}
\begin{eqnarray}
V_0^{(4)}={{6 M_{GUT}^6 e^{6kL}} \over\displaystyle 
M_{pl}^2} (1-e^{-2kL})
~,~\,
\end{eqnarray}  
\begin{eqnarray}
V_1^{(4)}=-{{6 M_{GUT}^6 e^{2kL}} \over\displaystyle 
M_{pl}^2} (1-e^{-2kL})
~.~\,
\end{eqnarray}  
Because $M_{GUT}$ can not be
smaller than TeV scale,
 each of above values is much larger than $10^{-58}$ $GeV^4$.

For above case,
the $M_1$ brane tension is negative (although it is not ruled out),
and the space is compact.
Now, we consider the case similar to that in the Ref. [3]. 
 There is just one positive tension
 brane
which is located at the origin and the fifth dimension is infinity. The
solution of the metric can be written as:
\begin{equation}
ds^2 = e^{- 2 k  |y| - 2 c} \eta_{\mu \nu} dx^{\mu} dx^{\nu} + dy^2
~,~
\end{equation} 
which is similar to the above case, and k is defined as before. The
scale relations are:
\begin{equation}
M_{pl}^2 =  {{M_X^3  e^{-2c}} \over\displaystyle k} 
 ~,~\,
\end{equation} 
\begin{equation}  
M_{GUT} = M_X e^{- c} ~,~ \,
\end{equation}
\begin{equation}
M_{pl}^2 =  {{M_{GUT}^3 e^{c}  }\over\displaystyle k} 
 ~.~\,
\end{equation} 
In this case, because the fifth dimension is not compactified,
we do not require that $k \simeq M_X$
~\footnote{ One may think there exists hierarchy between $k$ and $M_X$
in 5-dimensional fundamental metric which is not gauge hierarchy, but, 
k is related to the cosmology constant and
 small $k$ may ameliorate the cosmology constant 
 problem, which can be
 considered as the motivation of small $k$.}. 
We can push $M_{GUT}$ to TeV or $10^5$ GeV scale by
choosing $e^{-c}$ k=$10^{-27}$ GeV and
$10^{-21}$ GeV.
Taking c=0, $M_X$=$M_{GUT}$, and in general,
$M_X$ might be any number. 
In order not to be ruled out by experiment, we might
require that $ e^{-c} k > 10^{-13}$ GeV. Choosing c=0 and 
$k= 10^{-13}$ GeV,  we obtain  $M_{GUT}=M_X \simeq
10^8$ GeV, and $M_{GUT}/M_W \simeq M_{top}/M_{electron}$, so,
the hierarchy is not too severe.
However, we can not obtain small enough cosmology constant in this
scenario too, for the  cosmology constant and
the brane tension in the 4-dimensional Minkowski metric can be expressed as:
\begin{equation}
\Lambda^{(4)}= - {{3 M_{GUT}^6 }\over\displaystyle M_{pl}^2}
~,~ V_0^{(4)}= {{6 M_{GUT}^6 }\over\displaystyle M_{pl}^2}
 ~.~\,
\end{equation}

The results in Ref.[4] can also be applied here.
Assuming the brane at the orgin as a hidden brane, and our world locates at
$y_0$, we obtain the scale relations
\begin{equation}  
M_{H} = M_X e^{- c} ~,~ M_{GUT}=M_X e^{-k |y_0| - c} ~,~\,
\end{equation}
\begin{equation}
M_{pl}^2 =  {{M_{H}^3 e^{c}  }\over\displaystyle k} ~,~
M_{pl}^2 =  {{M_{GUT}^3 e^{3 k |y_0| + c}  }\over\displaystyle k} 
 ~,~\,
\end{equation} 
where the $M_H$ is the 4-dimensional
GUT scale on the hidden brane, and
we have assumed that the 5-dimensional GUT scale $M_H^{(5)}$ 
on the hidden brane is equal to the 5-dimensional Planck 
scale, i. e.,
$M_H^{(5)} = M_X$. One can easily solve the
gauge hierarchy problem in our world by varying $y_0$ or c.

In addition, in 4-dimensional Minkowski metric, the
cosmology constant $\Lambda^{(4)}$ is:
\begin{equation}
\Lambda^{(4)}= - {{3 M_{GUT}^6 e^{6 k|y_0|}}\over\displaystyle M_{pl}^2}
 ~.~\,
\end{equation}  
Therefore, the larger the $|y_0|$, the larger the magnitude of
the $\Lambda^{(4)}$ for fixed $M_{GUT}$. We can not explain
why the cosmology constant is so small, too.

In short, the general physical scale picture is the following: 
in 5-dimension,
 the GUT scale on each brane
is the same as the Planck scale, but, in 4-dimension,
the  Planck scale is generated from the our world low
GUT scale exponentially. The 4-dimensional GUT scales 
and Planck scale
are related to the 5-dimensional GUT scales and Planck scale 
by  exponential factors, respectively. 
In addition,  if the fifth dimension is not compact,
 there is a hidden brane with positive tension 
at the origin, the 4-dimensional GUT scale $M_H$ on the hidden
brane  is  larger than  or the same as 
that of our world.   we
live on a brane separated from the hidden 
 brane  a distance about $ln(M_H/M_{GUT})/k$ 
along the fifth dimension where $k$ is not necessary to 
be equal to the 5-dimensional Planck scale for the fifth 
dimension is not compact.

(II)  Assuming $\Lambda$ is negative, the other 
solution is:
\begin{equation}
\sigma =  -|y| \sqrt{{{- {\Lambda}} \over\displaystyle {6 M_X^3}}} + c
~.~\,
\end{equation}
Similar to the previously result, we
can define a new scale k which relates $V_0, V_1, \Lambda$ to $M_X$ as:
\begin{equation}
-V_{0} =  V_{1} = 6 M_X^3 k~,~ \Lambda = - 6 M_X^3 k^2
~.~\,
\end{equation} 
And then, the bulk metric is
\begin{equation}
ds^2 = e^{ 2 k  |y| - 2 c} \eta_{\mu \nu} dx^{\mu} dx^{\nu} + dy^2
~.~\,
\end{equation}

The Planck scale in the four-dimensional Minkowski metric is:
\begin{equation}
M_{pl}^2 =  {{M_X^3  e^{-2c}} \over\displaystyle k} [e^{2 k L}- 1]
 ~.~\,
\end{equation} 
We assume that the brane of our world is $M_0$. Using similar ansatz as
above, 
we obtain our world 4-dimensional GUT scale, and the relation between
$M_{pl}$ and $M_{GUT}$
\begin{equation}  
M_{GUT} = M_{GUT}^{(5)} e^{- c}= M_X e^{- c} ~,~ \,
\end{equation} 
\begin{equation}
M_{pl}^2 =  {{M_{GUT}^3  e^{2 k L + c}} \over\displaystyle k} [1- e^{-2 k L}]
 ~.~\,
\end{equation}
We can push $M_{GUT}$ to the TeV scale or $10^5$ GeV scale
easily. Taking $k=M_X$ for the fifth dimension is compact in 
this case, we obtain
\begin{equation}
M_{pl}^2 =  M_{GUT}^2  e^{2 k L } [1- e^{-2 k L}]
 ~.~\,
\end{equation}  
As before, with $k L= 34.5$ and 30, we can have  the $M_{GUT}$ at
the TeV scale  and $10^5$ GeV scale, respectively.
But, one should notice that, in this scenario, if we take c=0,
we will not rescale the mass scale in the Standard Model
from the 5-dimensional 
metric to the 4-dimensional metric, but we rescale the Planck scale,
the large 4-dimensional Planck scale is generated from the
low 5-dimensional 
Planck scale exponentially.

For c=0 and $k=M_X$, we obtain $M_X=M_{GUT}$. For
$c=kL/2$ and $k=M_X$, we obtain that for $M_{GUT} = 10^3$ GeV and $10^5$ GeV,
$M_X$ is about $10^{10}$ GeV and $ 10^{11}$ GeV, respectively.

One can also consider the cosmology constant and brane tensions in
the 4-dimensional Minkowski metric, they are:
\begin{eqnarray}
\Lambda^{(4)}=-{{3 M_{GUT}^6 e^{6kL}} \over\displaystyle 
M_{pl}^2} (1-e^{-2kL}) (1-e^{-4kL}) 
~,~\,
\end{eqnarray}
\begin{eqnarray}
V_0^{(4)}=-{{6 M_{GUT}^6 e^{2kL}} \over\displaystyle 
M_{pl}^2} (1-e^{-2kL})
~,~\,
\end{eqnarray}  
\begin{eqnarray}
V_1^{(4)}={{6 M_{GUT}^6 e^{6kL}} \over\displaystyle 
M_{pl}^2} (1-e^{-2kL})
~.~\,
\end{eqnarray}  
Obviosly, because $M_{GUT}$ can not be
smaller than TeV scale,
 each of those values is much larger than $10^{-58}$ $GeV^4$.  

Furthermore, this solution is related to the first solution by the
following trnsformation: $c \rightarrow c+ kL$ and
$V_0 \leftrightarrow V_1$. 

\section{Compactification on $AdS_5$ Space} 
In  all  the solutions of above section, the space is a slice of
$AdS_5$ space, therefore, we can consider the compactification on the
$AdS_5$ space. The $AdS_5$ metric is:
\begin{eqnarray}
ds^2= {r^2 \over\displaystyle  R_{ads}^2} \eta_{\mu \nu} dx^{\mu} dx^{\nu} +
{R_{ads}^2 \over\displaystyle  r^2} dr^2 ~,~\,
\end{eqnarray}
where $R_{ads}$ is the ``radius''of $AdS_5$ space, and 
$R_{ads}^2=\alpha' \sqrt{4 \pi gN} $ for Type IIB string on
$AdS_5 \times S^5$~\cite{JUAN} ( we do not write down the metric on $S^5$).
Obviously, there exists singularity when r is 0 or infinity. 

In order to be easy to compare with results of the previous  section,
 we make the following transformation
\begin{eqnarray}
r = R_{ads} e^{y/R_{ads}}
~.~\,
\end{eqnarray}
Then, the $AdS_5$ metric becomes:
\begin{equation}
ds^2 = e^{ 2 y/R_{ads} } \eta_{\mu \nu} dx^{\mu} dx^{\nu} + dy^2
~.~
\end{equation} 
This metric is invariant under the following transformation:
\begin{equation}
x^{\mu} \rightarrow e^{-\lambda/R_{ads}} x^{\mu}
~,~ y \rightarrow y + \lambda 
~.~
\end{equation} 
Assuming that the minimum and maximum of y are $a$ and b which can be
considered as cut-offs, and
the brane of our world is located at $y_o$ where $a \leq y_o \leq b$
~\footnote{ We do not address the issue of determining
the locations of the branes here.}, we obtain the
following scale relations: 
\begin{equation}
M_{pl}^2 =  {{M_X^3  R_{ads}} \over\displaystyle 2} e^{2b/R_{ads}}
(1-e^{-2(b-a)/R_{ads}})
 ~,~\,
\end{equation} 
\begin{equation}  
M_{GUT} = M_{GUT}^{(5)} e^{y_o/R_{ads}}= M_X e^{y_o/R_{ads}} ~,~ \,
\end{equation}
\begin{equation}
M_{pl}^2 =  {{M_{GUT}^3  R_{ads}} \over\displaystyle 2} e^{(2b-3 y_o)/R_{ads}}
(1-e^{-2(b-a)/R_{ads}})
 ~.~\,
\end{equation}
If $a$ and $b$ are finite, we can assume that $R_{ads}=M_X^{-1}$, and
obtain 
\begin{equation}
M_{pl}^2 =  {{M_{GUT}^2} \over\displaystyle 2} e^{2(b-y_o)/R_{ads}}
(1-e^{-2(b-a)/R_{ads}})
 ~.~\,
\end{equation}  
With $2(b-y_o)/R_{ads}$ =  34.5 and 30, we can push $M_{GUT}$ to
the TeV scale  and $10^5$ GeV scale, respectively. But, one should notice that,
if $y_o=0$, we will not rescale the mass scale in the Standard Model, which is
different from previous result~\cite{LRRS}.  If $a$ is negative infinity,
the physical scale picture will be similar to that in the first solution
with non-compact fifth dimension in above 
section, with $c=-b/R_{ads}$.

In above section, the first solution with compact fifth dimension
 can be considered as $a=y_o=b-L$ and $c=-b/R_{ads}$,
and the scond solution  can be considered as
$a=y_o=b-L$  and $c=-a/R_{ads}$ by noticing that $k=R_{ads}^{-1}$.
Of course, the value of $a$ might be negative infinity,
the result in the first solution with non-compact fifth
dimension can be considered as
$c=-b/R_{ads}$.

\section*{Acknowledgments}
We would like to thank V. Barger very much for reading the manuscript and
comments.  
We also would like to thank Yungui Gong and J. X. Lu for helpful discussion.
This research was supported in part by the U.S.~Department of Energy under
 Grant No.~DE-FG02-95ER40896 and in part by the University of Wisconsin 
 Research Committee with funds granted by the Wisconsin Alumni
  Research Foundation.

\end{document}